\documentclass[%
 reprint,
 amsmath,amssymb,
 aps,
prx
]{revtex4-1}

\usepackage{graphicx}
\usepackage{dcolumn}
\usepackage{bm}
\usepackage{epstopdf}
\usepackage{bbm}
\usepackage{bbold}
\usepackage[pdftex]{xcolor}

\begin{document}

\preprint{K233-HF}

\title{Temperature and Angular Dependence of the Upper Critical Field in K$_{2}$Cr$_{3}$As$_{3}$}

\author{Huakun Zuo$^{1,\#}$, Jin-Ke Bao$^{2,\#}$,  Yi Liu$^{2}$, Jinhua Wang$^{1}$, Zhao Jin$^{1}$, Zhengcai Xia$^{1}$, Liang Li$^{1}$, Zhuan Xu$^{2,3}$, Jian Kang$^{4}$, Zengwei Zhu$^{1, *}$ and Guang-Han Cao$^{2,3*}$}

\affiliation{1. Wuhan National High Magnetic Field Center, School of Physics, Huazhong University of Science and Technology,  Wuhan  430074, China\\
2. Department of Physics, Zhejiang University, Hangzhou 310027, China\\
3. Collaborative Innovation Centre of Advanced Microstructures, Nanjing 210093, China \\
4. School of Physics and Astronomy, University of Minnesota, Minneapolis,
Minnesota 55455, USA}

\date{\today}

\begin{abstract}
We report measurements of the upper critical field $H_{\mathrm{c2}}$ as functions of temperature $T$, polar angle $\theta$ (of the field direction with respect to the crystallographic $c$ axis), and azimuthal angle $\phi$ (of the field direction relative to the $a$ axis within the $ab$ plane) for the Cr-based superconductor K$_{2}$Cr$_{3}$As$_{3}$ with a quasi-one-dimensional and non-centrosymmetric crystal structure. We confirm that the anisotropy in $H_{\mathrm{c2}}(T)$ becomes inverse with decreasing temperature. At low temperatures, $H_{\mathrm{c2}}(\theta)$ data are featured by two maxima at $\theta$ = 0 ($\mathbf{H}\parallel c$) and $\pi/2$ ($\mathbf{H}\bot c$), which can be quantitatively understood only if uniaxial effective-mass anisotropy and absence of Pauli paramagnetic effect for $\mathbf{H}\bot c$ are taken simultaneously into consideration. The in-plane $H_{\mathrm{c2}}(\phi)$ profile shows a unique threefold modulation especially at low temperatures. Overall, the characteristic of the $H_{\mathrm{c2}}(\theta, \phi, T)$ data mostly resemble those of the heavy-fermion superconductor UPt$_3$, and we argue in favor of a dominant spin-triplet superconductivity with odd parity in K$_{2}$Cr$_{3}$As$_{3}$.
\\


\end{abstract}

\maketitle
\section{\label{sec:level1}Introduction}

Unconventional superconductors are those materials that possess exotic superconductivity (SC) whose origin cannot be explained by electron-phonon interactions in Bardeen-Cooper-Schrieffer (BCS) theory\cite{bcs}. A more operable description for unconventional superconductors addresses an additional symmetry broken\cite{sigrist}, apart from the $U(1)$-gauge symmetry and crystalline-lattice symmetry. Examples of unconventional superconductors include, chronologically, CeCu$_2$Si$_2$\cite{steglich}, organic superconductors\cite{osc}, UPt$_3$\cite{stewart}, high-$T_\mathrm{c}$ cuprates\cite{bm}, Sr$_2$RuO$_4$\cite{maeno}, UGe$_2$\cite{UGe2} and iron-based superconductors\cite{hosono}. Those novel superconductors bring rich interesting physics and challenge our understanding of SC\cite{norman}.

Recently, SC was discovered in a Cr-based family $A_2$Cr$_3$As$_3$ ($A$ = K\cite{bao}, Rb\cite{Rb233} and Cs\cite{Cs233}). The new materials possess a quasi-one-dimensional (quasi-1D) crystal structure characterized by infinite [(Cr$_3$As$_3$)$^{2-}$]$_{\infty}$ linear chains, called double-walled subnanotubes, which are separated by alkali-metal cations. The point group is $D_{3h}$, hence there is no inversion center for the crystal structure. The superconducting transition temperature $T_{\mathrm{c}}$ is 6.1 K, 4.8 K and 2.2 K, respectively, for $A$ = K, Rb and Cs. Unconventional SC in K$_2$Cr$_3$As$_3$ or Rb$_2$Cr$_3$As$_3$ has been supported by accumulating experimental and theoretical results as follows. (1) The Sommerfeld specific-heat coefficient is nearly 4 times of the value from the first-principles calculation\cite{caoc,hujp1}, indicating significant electron correlations in K$_2$Cr$_3$As$_3$. (2) K$_2$Cr$_3$As$_3$ shows a large upper critical field $H_{\mathrm{c2}}$, which exceeds the BCS weak-coupling Pauli limit\cite{clogston,chandrasekhar}, $H_{\mathrm{c2}}^{\mathrm{P}}$ = 18.4$T_{\mathrm{c}}\approx$ 110 kOe, by 3$-$4 times\cite{bao,canfield1,canfield2}. (3) The $^{75}$As nuclear quadrapole resonance (NQR) shows a strong enhancement of Cr-spin fluctuations above $T_{\mathrm{c}}$ and, there is no Hebel-Slichter coherence peak in the temperature dependence of nuclear spin-lattice relaxation rate just below $T_{\mathrm{c}}$ for K$_2$Cr$_3$As$_3$\cite{imai}. Similar result is given by the nuclear magnetic resonance (NMR) for Rb$_2$Cr$_3$As$_3$, from which \emph{ferromagnetic} spin fluctuations are additionally evidenced\cite{zgq}, supporting a spin-triplet pairing scenario. The latter seems to be consistent with the observation of a spontaneous internal magnetic field near $T_{\mathrm{c}}$, although being very weak, in the muon spin relaxation or rotation ($\mu$SR) experiment\cite{musr}. (4) Penetration-depth measurements indicate existence of line nodes in the superconducting gap\cite{yhq1}. (5) Band-structure calculations show that Cr-3$d$ orbitals dominate the electronic states at the Fermi level ($E_\mathrm{F}$) and, the consequent Fermi-surface sheets (FSs) consist of a three-dimensional (3D) FS in addition to two quasi-1D FSs\cite{caoc,hujp1,alemany}. Ferromagnetic and/or frustrated spin fluctuations are suggested by the calculations. (6) Theoretical models\cite{zy,hujp2,dai} are established based on the molecular orbitals, from which spin-triplet SC is stabilized. (7) The expected $T_\mathrm{c}$ suppression by impurity scattering for non-$s$-wave superconductors is observed in the K$_2$Cr$_3$As$_3$ crystals prepared using impure Cr\cite{Impurity}.

As is known, the $H_{\mathrm{c2}}$ behavior of a type-II superconductor may be an indicator for unconventional SC\cite{USC,PDwave,zhang}. The temperature and angular dependence of $H_{\mathrm{c2}}$ reflect the mechanisms of Cooper-pair breaking due to an orbital and/or Zeeman effect. $H_{\mathrm{c2}}(T)$ data of K$_2$Cr$_3$As$_3$ have been measured for different samples\cite{bao,canfield1,canfield2,wxf}. Measurements using single crystals revealed a large initial slope, $-$(d$H_{\mathrm{c2}}$/d$T)|$$_{T_{\mathrm{c}}}$, of 120 kOe/K\cite{canfield1} or 161 kOe/K\cite{wxf}, for field parallel to the crystallographic $c$ axis ($\mathbf{H}\parallel c$). $H_{\mathrm{c2}}^{\parallel}(T)$ exhibits a strongly negative curvature, and it saturates at about 230 kOe, indicating a Pauli-limiting scenario with significant spin-orbit coupling\cite{canfield2}. On the other hand, the $H_{\mathrm{c2}}^{\bot}(T)$ data basically show an orbitally limited behavior with no signs of paramagnetic pair breaking. Consequently, $H_{\mathrm{c2}}^{\bot}(T)$ and $H_{\mathrm{c2}}^{\parallel}(T)$ cross at $T\approx4$ K\cite{canfield2}.

To further understand the different behaviors of $H_{\mathrm{c2}}^{\parallel}(T)$ and $H_{\mathrm{c2}}^{\bot}(T)$, we measured the $H_{\mathrm{c2}}$ \textit{in situ} as functions of \emph{the polar angle $\theta$} (the angle relative to the $c$ axis), \emph{the azimuthal angle $\phi$} (the angle relative to the [10\={1}0] direction in the basal plane), as well as temperature $T$ for K$_{2}$Cr$_{3}$As$_{3}$ crystals. The extrapolated orbitally limited $H_{\mathrm{c2}}$ for $\mathbf{H}\parallel c$ and $\mathbf{H}\bot c$ at zero temperature exceed the Pauli limit by a factor of 4.6 and 3.4 respectively, far beyond the scope of singlet pairing scenario. The $H_{\mathrm{c2}}(\theta)$ data demonstrate that the apparent anisotropy reversal phenomenon is due to the paramagnetically pair-breaking effect \emph{only for the field component parallel to the $c$ axis}. The $H_{\mathrm{c2}}(\theta)$ completely satisfy the equation based on the assumption of triplet Cooper pairing. Furthermore, the $H_{\mathrm{c2}}(\phi)$ profile shows a unique threefold modulation, suggesting the coupling of a symmetry-breaking field. These results point to a dominant spin-triplet superconducting state in K$_{2}$Cr$_{3}$As$_{3}$.

\section{\label{sec:level2}Experimental Methods}
Rod-shape single crystals of K$_{2}$Cr$_{3}$As$_{3}$ were grown by a self-flux method\cite{bao}. The magnetoresistance was measured with a standard four-probe technique on a rotator under pulsed magnetic field which can reach 60 T. The four contacts were attached to 25 $\mu$m diameter gold wires by using Dupont 4929N silver paint [examples photographed in Figs. S4 and S7 of the Supplementary Materials (SM)]. The samples were mounted on the rotator probe with a sapphire substrate. All the procedures were done in an Ar-filled glove box. After transferring the probe from the glove box, we replaced Ar gas in the sample space with helium gas which was served as exchanging gas. The electrical current was applied along the rod direction which is always parallel to the crystallographic $c$ axis because of the crystallization habit. The rotational axis is either perpendicular or parallel to the rod. So the magnetic field direction could be adjusted \textit{in-situ}. The angles ($\theta$ and $\phi$) could be extracted from the ratio of a coil attached to the back of the rotating platform to the pick-up coil for field. Note that the sample was air-sensitive and couldn't be reused. We thus employed different samples for different measurements. Nevertheless, we checked that the results were reproducible.

\section{\label{sec:level3}Results and discussion}

\subsection{\label{subsec:level1}$H_{\mathrm{c2}}(T, \theta)$ Result}
At zero field, the sample shows a sharp superconducting transition at $T_\mathrm{c}$ = 6.2 K, as shown in Fig. S1 of SM. The residual resistance ratio (RRR), \emph{i. e.} a ratio of the resistivity at room temperature and at the temperature just above $T_\mathrm{c}$, is about 60, indicating high-quality of the sample. Note that a small residual resistance is left below $T_\mathrm{c}$. This is probably due to the chemical reaction between sample and the silver-paste electrodes, which forms a non-superconducting KCr$_{3}$As$_{3}$ phase\cite{K133}. Nevertheless, the partial deterioration of the sample does not affect the $H_{\mathrm{c2}}$ determination because the onset transition temperature $T_\mathrm{c}^{\mathrm{onset}}$ value is not altered, compared with the previous reports\cite{bao,canfield1,canfield2,wxf}.

Field dependence of the magnetoresistance was employed to determine the $H_{\mathrm{c2}}$ which is defined by the crossing point of the normal and superconducting states, as shown in Figs. 1a and 1b. The resulting $H_{\mathrm{c2}}$ data are plotted in Fig. 1c. Here we note that defining the $H_{\mathrm{c2}}$ as the peak position of the derivative of magnetoresistance does not change the result (see Figs. S2 and S3 in SM). Overall, the obtained $H_{\mathrm{c2}}-T$ phase diagram is similar to the previous profile\cite{canfield2}, featured by the crossing of $H_{\mathrm{c2}}^{\bot}(T)$ and $H_{\mathrm{c2}}^{\parallel}(T)$. The ``crossover" temperature is 3.1 K, somewhat lower than the value ($\approx$ 4 K) previously reported\cite{canfield2}. Apparently, the reversal in $H_{\mathrm{c2}}$ anisotropy suggests a dimension crossover from uniaxial anisotropy to planar one (note that there are "K$_{2}$Cr$_{3}$As$_{3}$" atomic planes at $z$ = 0 and $z$ = 1/2 in the crystal structure\cite{bao}). However, the $H_{\mathrm{c2}}(\theta)$ data in Fig.~\ref{fig2:theta} show modulations at the ``crossover" temperature of 3.1 K, which is inconsistent with the expected angle dependence, $H_{\mathrm{c2}}(\theta)\thicksim(\mathrm{cos}^{2}\theta+\epsilon^{2}\mathrm{sin}^{2}\theta)^{-1/2}$, from which a constant $H_{\mathrm{c2}}$ is concluded for the effective-mass ratio $\epsilon^{2} = 1$. Even at 1.9 K , where $H_{\mathrm{c2}}^{\bot}>H_{\mathrm{c2}}^{\parallel}$, a small peak at $\theta =$ 0$^{\circ}$ is present. This indicates that the uniaxial anisotropy is still there, and thus the dimension-crossover scenario can be ruled out.

\begin{figure*}
\includegraphics[width=15cm]{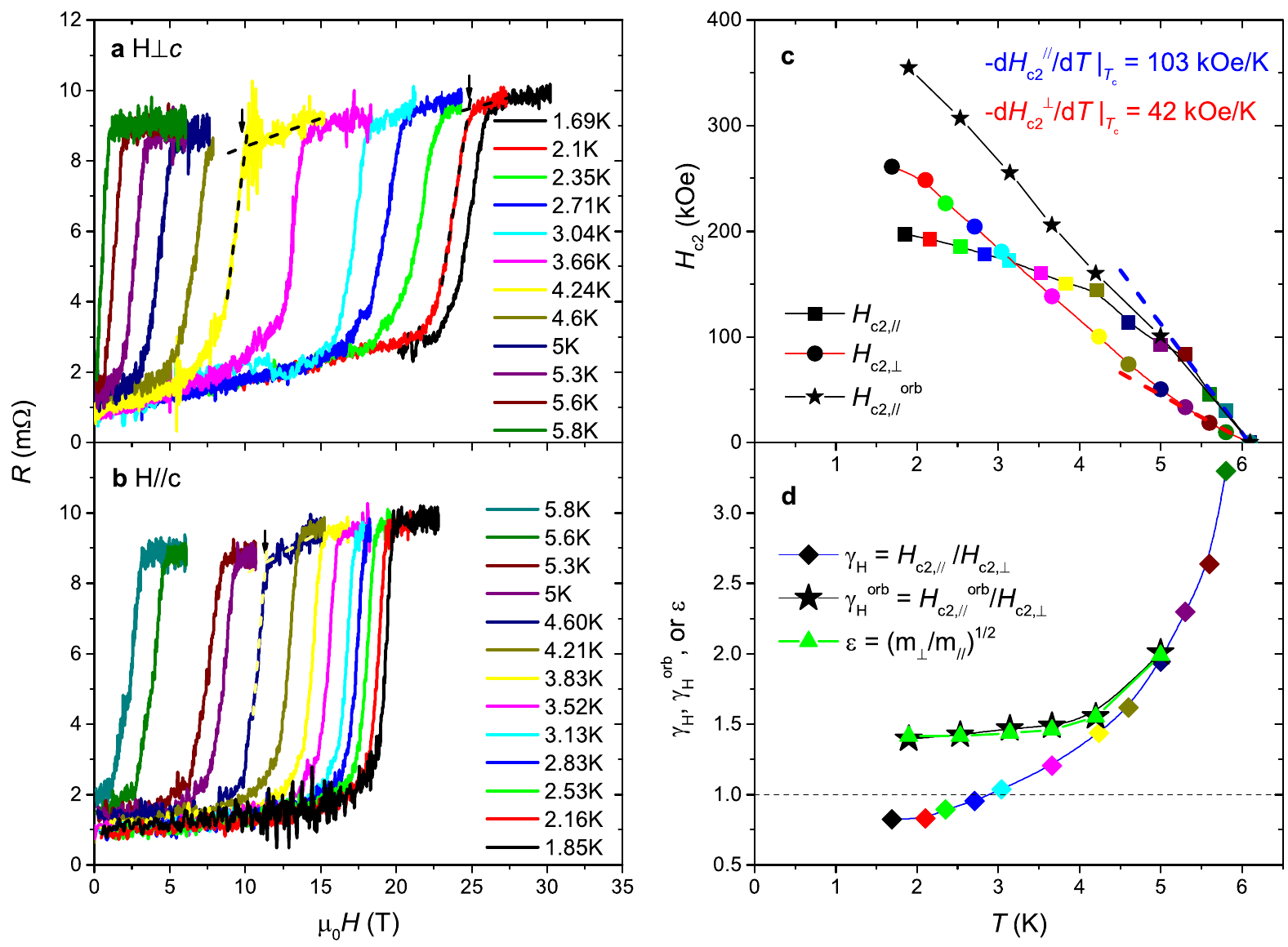}
\caption{\textbf{Determination of anisotropic $H_{\mathrm{c2}}$ for K$_{2}$Cr$_{3}$As$_{3}$ crystals.} \textbf{a} and \textbf{b}, Magnetoresistance as a function of field (\textbf{H}$\parallel c$ and \textbf{H}$\bot c$, respectively) at fixed temperatures. \textbf{c}, Temperature dependence of the anisotropic $H_{\mathrm{c2}}$, defined by the ``junction" point of superconducting and normal states shown in \textbf{a} and \textbf{b}. $H_{\mathrm{c2},\parallel}^{\mathrm{orb}}$ refers to the orbitally limited upper critical field for \textbf{H}$\parallel c$, which is obtained by the $H_{\mathrm{c2}}(\theta)$ data fitting shown in Fig. 2. \textbf{d}, The anisotropic parameters $\gamma_{H}(T) = H^{\parallel}_{\mathrm{c2}}/H^{\bot}_{\mathrm{c2}}$, $\gamma_{H}(T)$$^{\mathrm{orb}}$ = $H_{\mathrm{c2},\parallel}^{\mathrm{orb}}/H^{\bot}_{\mathrm{c2}}$, and $\epsilon$ (see definition in the text) as functions of temperature.}
\end{figure*}

The phenomenon of crossing of $H_{\mathrm{c2}}^{\bot}(T)$ and $H_{\mathrm{c2}}^{\parallel}(T)$, although being uncommon, was reported in several other systems including the quasi-1D organic superconductor (TMTSF)$_{2}$PF$_{6}$\cite{oscHc2}, iron-based chalcogenides\cite{FeTe1,FeTe2,FeTeS}, and the heavy-fermion superconductor UPt$_{3}$\cite{UPt3theta}. Nevertheless, details of the $H_{\mathrm{c2}}(T)$ crossing, which reflect the origin of the anisotropy reversal, differ from each other. In the organic superconductor (TMTSF)$_{2}$PF$_{6}$, $H_{\mathrm{c2}}(T)$ displays pronounced upward curvature without saturation for two field directions perpendicular to $\mathbf{c}^*$ at low temperatures\cite{oscHc2}. This feature can be explained in terms of dramatic reduction of orbitally pair-breaking effect because of a field-induced dimension crossover\cite{zhang}. As for iron-based superconductors, the anisotropy inversion appears accidentally (thus it is not ubiquitous) and, they tend to be isotropic (the anisotropy ratio approaches 1.0) at $T\rightarrow 0$. The $H_{\mathrm{c2}}(T)$ data reveal that Pauli-limiting effect is at work regardless of the field directions\cite{FeTe2,zocco}, consistent with spin-singlet Cooper pairing. In the case of UPt$_3$, by contrast, the reversed anisotropy is appreciable at zero temperature, and more importantly,  Pauli-limiting effect is absent for $\mathbf{H}\bot c $. This observation leads to a proposal of triplet SC with odd parity in UPt$_3$\cite{sauls}.

The case in K$_{2}$Cr$_{3}$As$_{3}$ mostly resembles that of UPt$_3$, yet it is qualitatively different from those of iron-based superconductors.  As can be seen in Fig. 1c, $H_{\mathrm{c2}}^{\parallel}(T)$ shows negative curvatures, and it saturates at lower temperatures, consistent with a dominant paramagnetically pair-breaking scenario. In contrast, $H_{\mathrm{c2}}^{\bot}(T)$ basically shows a linear behavior with a slightly positive curvature, or, a kink at about 5.5 K. Furthermore, below 3 K $H_{\mathrm{c2}}^{\bot}(T)$ values surpass $H_{\mathrm{c2}}^{\parallel}(T)$ significantly, achieving 270 kOe at 1.7 K, or 370 kOe at 0.6 K\cite{canfield2}, which are 2.5 and 3.4 times of the Pauli limit. These data clearly indicate that, for $H_{\mathrm{c2}}^{\bot}(T)$, paramagnetically pair-breaking effect is minor at most, and therefore, orbitally pair-breaking effect turn out to be dominant. Indeed, the data fit using Werthamer-Helfand-Hohenberg
theory\cite{whh} for a uniaxial, single-gap superconductor shows that the Pauli pair-breaking parameter is zero for $H_{\mathrm{c2}}^{\bot}(T)$\cite{canfield2}. Therefore, the crossing of $H_{\mathrm{c2}}^{\parallel}(T)$ and $H_{\mathrm{c2}}^{\bot}(T)$ in K$_{2}$Cr$_{3}$As$_{3}$ comes from different pair-breaking mechanisms for the different external field directions, an extremely anisotropic Pauli-limiting effect, in contrast with the dominant Pauli-limiting scenario in iron-based superconductors\cite{FeTe2,zocco}.

The anisotropic Pauli-limiting character manifests itself in the $H_{\mathrm{c2}}(\theta)$ data, as shown in Fig. 2. Nearby $T_\mathrm{c}$, $H_{\mathrm{c2}}$ decreases monotonously as the field direction is tilted from $\mathbf{H} \| c$. This can be understood in terms of a uniaxial effective-mass anisotropy in Ginzburg-Landau (GL) theory. As the temperature is decreased, an additional maximum in $H_{\mathrm{c2}}(\theta)$ appears at $\theta = \pi/2$, consistent with the \emph{absence} of Pauli-limiting effect for $\mathbf{H}\bot c$. Below we show that the whole $H_{\mathrm{c2}}(\theta)$ data set can be perfectly explained by the combination of a \emph{fully} anisotropic Pauli-limiting effect with a uniaxial effective-mass anisotropy.

\begin{figure}
\includegraphics[width=8cm]{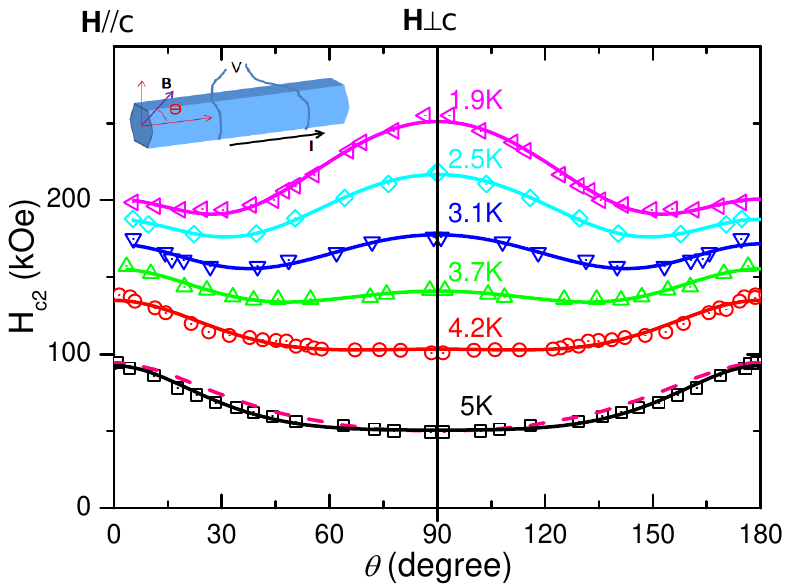}
\caption{\textbf{The out-of-plane $H_{\mathrm{c2}}(\theta)$ at various temperatures for K$_{2}$Cr$_{3}$As$_{3}$.} The solid lines and the dashed pink curve (for the 5 K data only) are the fitted data using Eq.~\ref{Eqn:Hc2} and Eq.~\ref{Eqn:orb}, respectively.}
\label{fig2:theta}
\end{figure}

First of all, according to GL theory, the effective-mass anisotropy leads to the anisotropy of the orbitally limited $H_{\mathrm{c2}}^{\mathrm{orb}}(\theta)$,
\begin{equation}
H_{\mathrm{c2}}^{\mathrm{orb}}(\theta)=\frac{H_{\mathrm{c2},\parallel}^{\mathrm{orb}}}{\sqrt{\mathrm{cos}^{2}\theta+\epsilon^{2}\mathrm{sin}^{2}\theta}},
\label{Eqn:orb}
\end{equation}
where $H_{\mathrm{c2},\parallel}^{\mathrm{orb}}$ denotes the presumed orbitally limited upper critical field for \textbf{H}$\parallel c$ and, $\epsilon^{2}=m_{\bot}/m_{\parallel}=(H_{\mathrm{c2},\parallel}^{\mathrm{orb}}/H_{\mathrm{c2},\bot}^{\mathrm{orb}})^2$ is the effective-mass ratio. Indeed, Eq.~\ref{Eqn:orb} can basically describe the experimental $H_{\mathrm{c2}}(\theta)$ data at 5 K (yet with obvious deviations around $\theta$ = 30 $^{\circ}$). However, it completely fails to catch the $H_{\mathrm{c2}}(\theta)$ data at lower temperatures.

So we have to include the paramagnetic pair breaking due to Zeeman effect on the superconducting Cooper pairs, which is assumed to be fully anisotropic. The paramagnetically pair-breaking effect can be parameterized by an effective Pauli-limiting field $H_{\mathrm{pm}}$. Since only the magnetic-field component parallel to $c$ axis further suppresses the $H_{\mathrm{c2}}^{\mathrm{orb}}(\theta)$ in Eq.~\ref{Eqn:orb}, $H_{\mathrm{pm}}$ is then given by $H_{\mathrm{\mathrm{pm}}}(\theta) = H_{\mathrm{pm}}^{\parallel}\mathrm{cos}\theta$,
such that $H_{\mathrm{\mathrm{pm}}}(\theta) = H_{\mathrm{pm}}^{\parallel}$ (full Pauli limiting) for $\theta$ = 0 and, $H_{\mathrm{\mathrm{pm}}}(\theta)$ = 0 (absence of Pauli limiting) for $\theta = \pi/2$.

The effective Pauli-limiting field $H_{\mathrm{pm}}$ suppresses $H_{\mathrm{c2}}^{\mathrm{orb}}$ in a way of competition between the related energies\cite{clogston,chandrasekhar,whh,hake}. Given $E\sim H^2$, therefore,
$[H_{\mathrm{c2}}(\theta)]^2=[H_{\mathrm{c2}}^{\mathrm{orb}}(\theta)]^2-[H_{\mathrm{\mathrm{pm}}}(\theta)]^2.
$
Consequently, one obtains an explicit expression for $H_{\mathrm{c2}}(\theta)$,
\begin{equation}
H_{\mathrm{c2}}(\theta)=\sqrt{\frac{(H_{\mathrm{c2},\parallel}^{\mathrm{orb}})^2}{\mathrm{cos}^{2}\theta+\epsilon^{2}\mathrm{sin}^{2}\theta}-(H_{\mathrm{pm}}^{\parallel}\mathrm{cos}\theta)^2}.
\label{Eqn:Hc2}
\end{equation}

Remarkably, the whole experimental data set of $H_{\mathrm{c2}}(\theta, T)$ satisfies the above equation very well, as shown in Fig.~\ref{fig2:theta}, which further confirms the absence of Pauli-limiting effect for $\mathbf{H}\bot c$. The data fitting yields three parameters, $H_{\mathrm{c2},\parallel}^{\mathrm{orb}}$, $\epsilon$ and $H_{\mathrm{pm}}^{\parallel}$ at a given temperature. As expected, the fitted $H_{\mathrm{c2},\parallel}^{\mathrm{orb}}(T)$ is basically linear, as shown in Fig. 1c. The linear extrapolation yields a zero-temperature orbitally limited $H_{\mathrm{c2},\parallel}^{\mathrm{orb}}(0)$ of 515 kOe for $\mathbf{H}\parallel c$, which is reasonably larger than the extrapolated value of $H_{\mathrm{c2},\bot}^{\mathrm{orb}}(0)$ = 372 kOe for $\mathbf{H}\bot c$ (experimentally, $H_{\mathrm{c2},\bot}^{\mathrm{orb}}$ = 370 kOe at 0.6 K\cite{canfield2}). The obtained $H_{\mathrm{c2},\parallel}^{\mathrm{orb}}(0)$ and $H_{\mathrm{c2},\bot}^{\mathrm{orb}}(0)$ exceed the Pauli limit $H_{\mathrm{c2}}^{\mathrm{P}}$ by a factor of 4.6 and 3.4, respectively. Using the GL relations, $H_{\mathrm{c2},\parallel}^{\mathrm{orb}}(0) = \Phi_{0}/[2\pi \xi_{\bot}(0)^2$] and $H_{\mathrm{c2},\bot}^{\mathrm{orb}}(0) = \Phi_{0}/[2\pi \xi_{\bot}(0) \xi_{\parallel}(0)$], where $\Phi_{0}$ is the flux quantum, the anisotropic coherence lengths at zero temperature can be estimated to be $\xi_{\bot}(0)$ = 2.53 nm and $\xi_{\parallel}(0)$ = 3.50 nm. The $\xi_{\bot}(0)$ value exceeds twice of the interchain distance in K$_{2}$Cr$_{3}$As$_{3}$\cite{bao}, indicating a uniaxially anisotropic 3D SC.


Fig. 1d plots the anisotropic parameters, $\gamma_{H}(T)$, $\gamma_{H}(T)$$^{\mathrm{orb}}$ [=$H_{\mathrm{c2},\parallel}^{\mathrm{orb}}(T)/H^{\bot}_{\mathrm{c2}}(T)$] and $\epsilon(T)$. They tend to emerge at about 5 K. $\gamma_{H}(T)$ shows a divergence behavior at temperatures close to $T_\mathrm{c}$, implying the relevance of quasi-1D scenario to the emergence of SC. At 5.8 K, $\gamma_{H}$ is 3.3, corresponding to a effective-mass ratio of $m_{\bot}/m_{\parallel}\sim$ 11, which virtually reflects the obviously uniaxial anisotropy due to the quasi-1D crystal and electronic structures.  Upon cooling down, $\gamma_{H}(T)$ decreases rapidly and, it crosses the isotropic line of $\gamma_{H}(T)$ = 1.0 which was referred to as an ``anisotropy reversal"\cite{canfield2}. However, $\gamma_{H}(T)$$^{\mathrm{orb}}$ and $\epsilon(T)$ do not cross the $\gamma_{H}(T)$ = 1.0 line, which means that the anisotropy reversal would \emph{disappears} if the anisotropic Pauli-limiting effect were not involved. Indeed, $\gamma_{H}(T)$$^{\mathrm{orb}}$ and $\epsilon(T)$ saturate at 1.4 down to lower temperatures, consistent with the 3D SC concluded above.

\subsection{\label{subsec:level2}$H_{\mathrm{c2}}(\phi)$ Result}

The similarity of $H_{\mathrm{c2}}(T, \theta)$ between K$_{2}$Cr$_{3}$As$_{3}$ and UPt$_3$ motivates us to measure $H_{\mathrm{c2}}$ as a function of the azimuthal angle $\phi$, since the latter superconductor exhibits a sixfold modulation in the in-plane $H_{\mathrm{c2}}(\phi)$\cite{UPt3phi}. Fig. 3 shows derivative of the magnetoresistance for the field directions with different $\phi$ angles, from which the $H_{\mathrm{c2}}(\phi)$ can be determined by the peak value in d$R$/d$H$ (the peak values and their error bars were extracted by a Gaussian fit). This definition is consistent with the former one for the out-of-plane $H_{\mathrm{c2}}(\theta)$, which is shown in SM. The $\phi$ angle varied from $1.87^{\circ}$ to $153.38^{\circ}$. To plot a complete polar diagram, the data were rotated by $120^{\circ}$ and $-120^{\circ}$, respectively, according to the crystal symmetry with the point group of $D_{3h}$. The overlapped region validates the rotation. Consequently, the obtained $H_{\mathrm{c2}}(\phi)$ profile shows a three-fold (quasi-six-fold) modulation with an amplitude of 3.6 kOe. The maximum of $H_{\mathrm{c2}}(\phi)$ appears for the field directions along the crystallographic $a$ and $b$ axes. We can exclude the possibility of surface superconductivity and the influence of anisotropic normal-state magnetoresistance (see SM for the details). At a higher temperature of 4.23 K, the in-plane anisotropy magnitude decreases obviously (0.5 kOe), nevertheless, similar threefold modulation is still observable with the maximum basically along the $a$ axis, as shown in Figs. S6 and S9 of SM. Note that the $H_{\mathrm{c2}}(\phi)$ modulation (with a relative magnitude up to $\sim$1.2\%) in K$_{2}$Cr$_{3}$As$_{3}$ is actually more obvious than those in UPt$_3$\cite{UPt3phi} and MgB$_2$\cite{MgB2}, both of which also crystallize in a hexagonal lattice.
\begin{figure}
\includegraphics[width=8cm]{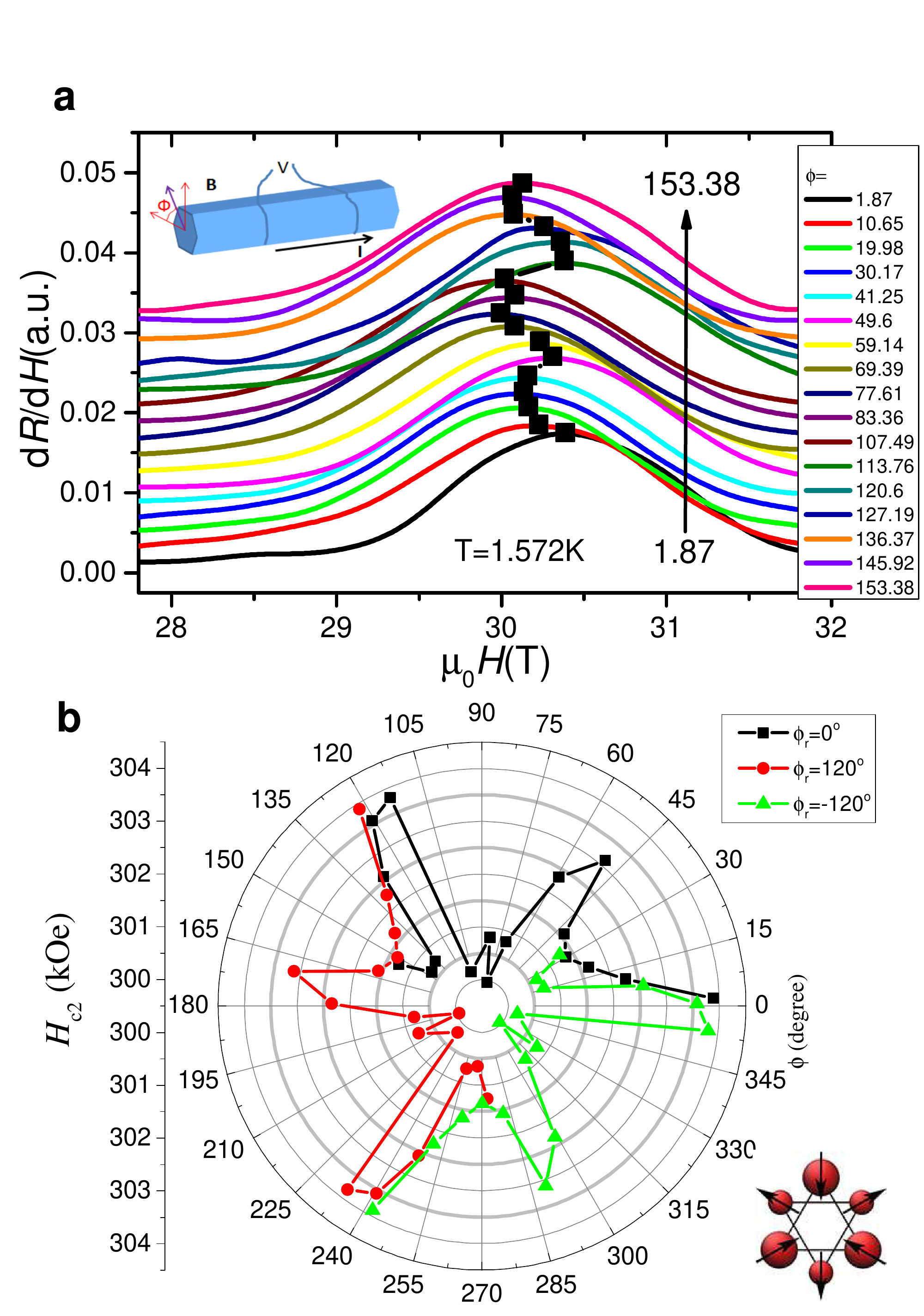}
\caption{\textbf{In-plane $H_{\mathrm{c2}}(\phi)$ at 1.572$\pm$0.001 K for K$_{2}$Cr$_{3}$As$_{3}$.} \textbf{a}. The derivatives of magnetoresistance $R(H)$ for different $\phi$ angles, from which $H_{\mathrm{c2}}$ is extracted. \textbf{b}. Polar plot of the extracted $H_{\mathrm{c2}}(\phi)$. The red (blue) triangles come from the rotation of the derived data with $\phi_r$ = 120$^{\circ}$ ($-120^{\circ}$), respectively. Shown at the lower-right corner is the possible Cr$_6$-octohedron spin configuration for the magnetic ground state, according to first-principles calculations.}
\end{figure}

The sixfold modulation of $H_{\mathrm{c2}}(\phi)$ in UPt$_3$ can be explained in terms of a coupling to the symmetry breaking field\cite{UPt$_{3}$Theory1}. Neutron diffraction study for UPt$_3$ indeed shows an antiferromagnetism whose moments are lying within the basal plane\cite{aeppli}. In the case of K$_{2}$Cr$_{3}$As$_{3}$, according to the first-principles calculations, it is nearby an in-out co-planar magnetic ground state\cite{hujp1,caoc}, shown at the lower-right corner in Fig. 3b, in which the magnetic moments lie in the basal plane. This result seems to be compatible with the observed $H_{\mathrm{c2}}(\phi)$ modulation that can be similarly explained by the coupling to the symmetry breaking field\cite{UPt$_{3}$Theory1}. Given the relatively small magnitude which tends to decrease when approaching $T_\mathrm{c}$, alternatively, the $H_{\mathrm{c2}}(\phi)$ anisotropy may also reflect the symmetry of crystalline lattice. It was shown that weak in-plane modulations of $H_{\mathrm{c2}}(\phi)$ can be resulted from the GL theory incorporating higher-order gradient terms\cite{ref1,ref2}, particularly for a periodic array of weakly-coupled superconducting filaments\cite{ref3} which seems to be relevant to K$_{2}$Cr$_{3}$As$_{3}$.

\subsection{\label{subsec:level3}Discussion}

For a conventional superconductor with a high $H_{\mathrm{c2}}(0)$ comparable to $H_{\mathrm{c2}}^{\mathrm{P}}$, the $H_{\mathrm{c2}}(0)$ value is normally limited by paramagnetic effect regardless of field directions. Considering the paramagnetically limited behavior of $H_{\mathrm{c2}}^{\parallel}(T)$ in K$_{2}$Cr$_{3}$As$_{3}$, as well as the preliminary observation of insensitivity of $T_\mathrm{c}$ on impurity scattering, Balakirev at al.\cite{canfield2} proposed a novel spin-singlet superconductivity. The absence of Pauli-limiting effect for \textbf{H}$\bot c$ is explained by assuming electron-spin locking along the $c$ direction.

The absence of Pauli-limiting effect for $H_{\mathrm{c2}}^{\bot}(T)$ and a large $H_{\mathrm{c2}}^{\bot}(0)$ value (3.4 times of $H_{\mathrm{c2}}^{\mathrm{P}}$) also dictate a spin-triplet pairing scenario, as in the case of UPt$_3$ (see Table I of SM for the comparison) that was recently confirmed to host a spin-triplet odd-parity SC\cite{E1u,schemm}. The (pseudo)spins of the odd-parity Cooper pairs are $\mid \uparrow\downarrow>+\mid \downarrow\uparrow>$ with $S_z$ = 0, which is equivalent to the spin state of $\mid \leftleftarrows>+\mid \rightrightarrows>$. In this circumstance, the Zeeman energy breaks the Cooper pairs for $\mathbf{H}\|c$, hence the Pauli-limiting behavior for $H_{\mathrm{c2}}^{\parallel}(T)$. By contrast, the perpendicular field simply changes the population of Cooper pairs with spin directions $\mid \leftleftarrows>$ and $\mid \rightrightarrows>$, and therefore, no paramagnetic pair-breaking is expected for $H_{\mathrm{c2}}^{\bot}(T)$. Indeed, by taking spin-triplet Cooper pairs into consideration, and with some simplifications and approximations, we are able to derive an equation for $H_{\mathrm{c2}}(\theta)$, whose solution is consistent with Eq.~\ref{Eqn:Hc2} (see SM for details). Here we note that, owing to the non-centrosymmetric crystal structure in K$_{2}$Cr$_{3}$As$_{3}$, the pairing symmetry is in principle a mixture of singlet and triplet states\cite{zy}, except for the case of simple $p_z$ wave in which a purely triplet pairing is anticipated because of the mirror-plane reflection symmetry\cite{hujp2}.

Finally, we comment on the impurity scattering effect on $T_\mathrm{c}$, which is important to judge the possibility of either singlet or triplet pairing state. For an odd-parity unconventional superconductor, nonmagnetic scattering serves as a source of pair breaking even at zero field, hence $T_\mathrm{c}$ suppression is expected. Note here is that, however, such an effect will not be evident in the clean-limit regime, i.e. the electron mean free path $l$ is much larger than the superconducting coherence length $\xi$. In K$_{2}$Cr$_{3}$As$_{3}$, $l$ is estimated to be $\sim$75 nm (after the electron-mass renormalization is considered) for the electron transport along the $c$ axis in the sample with RRR = 61 (see the SM), and $\xi_{\parallel}(0)$ is only 3.5 nm, thus $l\gg \xi$ holds. This explains why the $T_\mathrm{c}$ keeps almost unchanged for single-crystal samples with different RRRs. Our recent study shows that, when introducing sufficient impurities ($l\sim \xi$), $T_\mathrm{c}$ is indeed suppressed in K$_{2}$Cr$_{3}$As$_{3}$\cite{Impurity}.

\section{\label{sec:level4}Conclusion}

In conclusion, we have performed detailed $H_{\mathrm{c2}}(\theta,\phi,T)$  measurements on superconducting K$_{2}$Cr$_{3}$As$_{3}$ crystals. We confirm the ``anisotropy-reversal" phenomenon which reflects different pair-breaking mechanisms for different magnetic-field directions. The absence of paramagnetic pair breaking for $\mathbf{H}\bot c$ is further demonstrated by the $H_{\mathrm{c2}}(\theta)$ data set. The extracted values of $H_{\mathrm{c2},\parallel}^{\mathrm{orb}}(0)$ and $H_{\mathrm{c2},\bot}^{\mathrm{orb}}(0)$ (515 and 372 kOe, respectively) far exceed the Pauli limit. We also observe a three-fold modulation in $H_{\mathrm{c2}}(\phi)$. While the spin structure of the superconducting Cooper pairs cannot be definitely determined by the results above, spin-locking scenario should be essential. We argue that a dominant spin-triplet pairing is more natural to meet the experimental data. Note that the spin-triplet pairing is consistent with the previous implications/suggestions in experimental and theoretical studies\cite{caoc,hujp1,imai,zgq,musr,yhq1,zy,hujp2,dai}. We hope our result helps to find the exact superconducting order parameter in K$_{2}$Cr$_{3}$As$_{3}$, which might expand the overall understanding of unconventional superconductivity.

\section* {Achnowledgements}
We thank D. F. Agterberg, F. C. Zhang, Y. Zhou, C. Cao, J. H. Dai, F. Yang, X. X. Wu, R Fernandes, J. P. Hu and Kamran Behnia for helpful discussions. J.~K. is supported  by  the  U.S. Department  of  Energy,  Office  of  Science, Basic  Energy Sciences, under Award number de-sc0012336. Z. Z. is supported by the 1000 Youth Talents Plan, the National Science Foundation of China (Grant No. 11574097) and The National Key Research and Development Program of China (Grant No.2016YFA0401704). G. C. is supported by National Key R \& D Program of the MOST of China (Grant No. 2016YFA0300202) and the National Science Foundation of China (Grant No. 11674281).

\#These authors equally contributed to this work.\\
* \verb|zengwei.zhu@hust.edu.cn|\\
* \verb|ghcao@zju.edu.cn|\\

\end{document}